\documentclass[prl,twocolumn,aps]{revtex4-1}
\usepackage{amssymb}
\usepackage{amsmath}
\usepackage{graphicx}
\usepackage{dcolumn}
\usepackage{bm}
\begin{document}
\title{Feasibility of the optical fiber clock}

\author{Ekaterina Ilinova$^1$}

\author{James F. Babb$^2$ }

\author{Andrei Derevianko$^1$}

\affiliation{$^1$Department of Physics, University of Nevada, Reno,
Nevada 89557, USA}

\affiliation{$^2$Harvard-Smithsonian Center for Astrophysics, 60 Garden St., MS 14, Cambridge, Massachusetts 02138, USA}

\begin{abstract}
We explore the feasibility of a fiber clock, a compact high-precision optical lattice atomic clock based on atoms trapped inside hollow core
optical fiber. Such setup offers an intriguing potential for both substantially increased number of interrogated atoms and miniaturization. We evaluate the sensitivity of the $^1S_0$-$^3P_0$ clock transition in Hg and other divalent atoms to the fiber inner core surface at non-zero temperatures. The Casimir-Polder interaction induced $^1S_0$-$^3P_0$  transition frequency shift is calculated for the atom inside the hollow capillary as a function of atomic position, capillary material, and geometric parameters.  For $\mathrm{Hg}$ atoms on the axis of a silica capillary with inner radius $\geq 15 \,\mu \mathrm{m}$ and optimally chosen thickness $d\sim 1 \,\mu \mathrm{m}$, the atom-surface interaction induced  $^1S_0$-$^3P_0$ clock transition frequency shift  can be kept on the level $\delta\nu/\nu_{\mathrm{Hg}} \sim10^{-19}$.
We also estimate the atom loss and heating due to
the collisions with the buffer gas, lattice intensity noise induced heating, spontaneous photon scattering, and residual birefringence induced frequency shifts.
\end{abstract}

\maketitle

\section{Introduction}
\begin{figure}[h]
\begin{center}
\includegraphics*[width=3.3in]{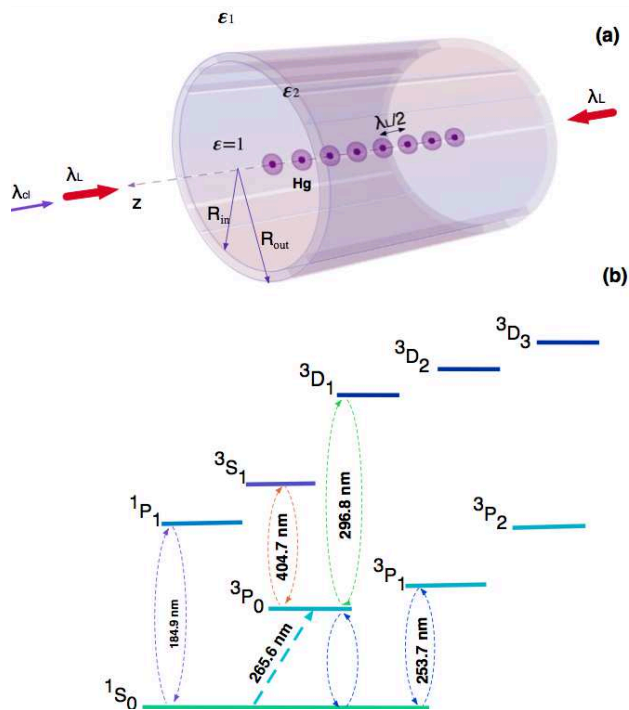}
\end{center}
\caption
{(a) Cold atoms optically trapped in the ground vibrational state inside the hollow core dielectric cylinder.  (b) Energy level diagram for Hg atoms. The clock transition is shown with the dashed straight arrow.  Dashed lines show the virtual transitions contributing to the resonant part of the Casimir-Polder interaction potential. 
\label{Fig:Setup}}
\end{figure}

The rapid progress in micro- and nano-fabrication technologies and advances in material science  together with the growing demand  for precision metrology and quantum information processing gave rise to  a new branch of research engineering aiming to develop  miniaturized quantum devices. Examples include atomic clocks~\cite{TreHomSte04,RamLacRos11, DeuRamLac10,GerWhiFer07}, atomic sensors, interferometers \cite{GriKnaKit10,LusYou08,AbbKanPat07,EklShkKna08,WanAndDan05}, and  quantum logic gates~\cite{Shu09}. In particular, development of miniaturized atomic clocks is anticipated to enable  important applications requiring portability and low power consumption,
such as secure telecommunications~\cite{FroYua15,Car13}, mobile timing, 
navigation~\cite{Lan91,Kit07,KraSho16},  and deep space atomic clocks~\cite{ElyMurSeu14}.

The main platforms  for compact quantum devices are micro-electromechanical systems~\cite{Kna08,KisHacFuj06}, miniaturized wire 
traps~\cite{ForGroZim98}, and, more recently, hollow core optical fibers~\cite{CreManKni99, OkaTakBen14}. Here we  analyze the feasibility of a ``fiber clock''---a device that holds interrogated clock atoms inside a hollow core optical fiber. The atoms are optically trapped, avoiding collisions with the wall surface~\cite{OkaTakBen14}.  
The ``fiber clock'' is a natural extension of optical lattice clocks~\cite{KatoriProc02}, which recently reached record levels of estimated fractional inaccuracy at the $\approx6\times 10^{-18}$ level~\cite{BloNicWil14}. The clock transition is  the narrow   $^1S_0$-$^3P_0$ transition present in divalent atoms (e.g. Sr, Yb, Ca, Mg, and Hg).  
Unlike in the free space configuration, where the maximum interaction length does not exceed the Rayleigh length $z_R=\pi w_0^2/\lambda$ (here $w_0$ is the characteristic Gaussian beam waist radius and $\lambda$ is the wavelength of the laser field), the fiber clock setup does not suffer from this limitation \cite{OkaTakBen14}.  Large ensembles of cold atoms trapped in a 1D optical lattice can be thus realized within the compact transverse region $\sim 5-100$ $\mu m$, avoiding high atomic density per lattice site. The increase in the number of interrogated atoms enables further advances in clock stability. Improved stability, i.e. the ability to average down statistical noise in shorter time intervals, is beneficial in many applications. In particular, searches for short transient variation of fundamental constants induced by ``clumpy'' dark matter~\cite{DerPos14,WkiMorBob16} would benefit from the improved clock stability.

In pursuing a super-precise miniature clock one also needs to keep in mind miniaturization of other clock elements.  In this context it is important to note that the  ``fiber clock''  can be  integrated into the all-fiber framework. 
Moreover one could envision the enhanced superradiance \cite{GroHar82} of 1D trapped atomic ensemble into the fiber guided laser mode. Optical trapping of large ensembles of cold divalent atoms uniformly distributed over the length  $L\sim N_a\lambda_{\mathrm{L}}/2$  (where $N_a\gg1$ is the number of trapped atoms and $\lambda_{\mathrm{L}}$ is the lattice wavelength) inside the hollow core fiber and suppressing the associated sources of atomic dipole relaxation (other than the radiative decay) will be the next step towards an ultra-narrow linewidth radiation source at  the $^1S_0$-$^3P_0$ clock transition frequency.   Super-radiant lasing on the clock transition \cite{MeiYeCar09,YuJin07} is a potential alternative to space-consuming bulky reference cavities used in optical clocks.

Here we theoretically evaluate the feasibility of building the optical lattice clock based on the narrow  $^1S_0$-$^3P_0$ transition in Hg and other alkaline-earth-like atoms (Cd,  Mg, Yb, Sr) optically trapped inside a hollow core fiber.  One of the first  experimental efforts towards compact optical clocks  involved the 3D trapping of an ensemble of Sr atoms  in a micron-sized structure~\cite{KisHacFuj06}. More recently precision spectroscopy of  the $^1S_0$-$^3P_1$ transition of Sr atoms optically trapped inside the Kagome fiber has been performed \cite{OkaTakBen14}.  
To the best of our knowledge no $^1S_0$-$^3P_0$  transition based optical lattice clock has been realized with cold atoms inside a hollow core fiber.  A mercury clock \cite{HacMiyPor08,PetChiDaw08,  McFMejMan12,TakUshDas15,TyuFavBil16} is of particular interest in applications to probing  physics beyond the standard model due to the large value of the Hg nuclear charge. 
The relatively low static polarizability of Hg and Cd \cite{HacMiyPor08} makes them least sensitive to black body radiation (BBR), as compared to the other atoms Sr, Yb, Ca. This pecularity makes Hg and Cd good candidates for optical clock applications where the ambient temperature is difficult to control, making the BBR effects  main constraint on the clock accuracy \cite{PorDer06}. 

 As a platform for the Hg and Cd clocks, one could consider recently developed  UV guiding hollow core fibers  \cite{FevGerLab09,GerFroWei14}. 
 In  \cite{GerFroWei14} a single mode photonic crystal fiber with inner core $\approx 20\, \mathrm{\mu}m$  and loss  $\approx 0.8\, \mathrm{dB/m}$  at the wavelength 280 nm was demonstrated. In \cite{FevGerLab09} a multi-mode hollow core fiber with  inner core diameter $\approx 25\, \mathrm{\mu}m$ and  loss $\approx 2\, \mathrm{dB/m}$ at wavelength 355 nm and loss of $\approx 0.4\, \mathrm{dB/m}$  at the wavelength 250 nm was  designed. 
 The ability to design similar waveguides for the UV range has been announced \cite{FevGerLab09}.

The development of a fiber-based atomic clock requires detailed understanding of the effects of the surrounding surface on the clock transition frequency.
The previous evaluation of the surface-induced clock frequency shifts for
divalent atoms was carried out for planar geometry~\cite{DerObrDzu09}.
Here we take into account the cylindrical fiber geometry and its material properties.
The atom-surface  interaction generally depends on the geometry and material of the fiber.  
 In \cite{EliBuhSch10} the Casimir-Polder (CP) interaction of Rydberg atoms with a cylindrical cavity was analyzed. It was shown that at certain cavity radii   an enhancement of modes resonant with atomic transitions may occur, leading to an increase of the resonant part of CP interaction potential and to the modification of the atomic  radiative decay rate~\cite{EliBuhSch10, JheAndHin87}.  

Considering the high precision required for the optical clock, we  study these resonant effects on the ground  $^1S_0$ and metastable $^3P_0$ clock states of alkaline-earth-like atoms. We present  the general form of the long range atom-surface interaction potential at non-zero temperatures for the hollow core cylindrical geometry and analyze the resulting $^1S_0$-$^3P_0$ clock frequency shift as a function of   the surface interface parameters.  
We find that for an ensemble of $\mathrm{Hg}$ atoms optically trapped near the axis of a hollow silica capillary waveguide with inner radius  $R_{in}\geq 15\; \mu \mathrm{m}$, the CP interaction-induced fractional frequency shift can be suppressed down to the level of $\delta\nu/\nu\sim10^{-19}$. The frequency shift due to the nonresonant part of the CP interaction in this case is dominant compared to the resonant contribution.  For the atoms trapped near the capillary axis the CP interaction decreases with the growth of the inner core radius. At high relative permittivity of the inner core surface material $\varepsilon_r\gg1$ the contribution of resonant atom-surface interaction as well as the radiative decay rate enhancement (Purcell effect) may become dominant at certain choices of the geometric parameters of the waveguide.  
Both the clock transition frequency shift caused by the resonant part of CP interaction potential and the Purcell effect can be suppressed  by slight adjustment of the thickness $d$ at given inner core radius $R_{in}$. The adjustment has to be done in order to avoid  the resonant waveguide modes at the frequencies of the decay channels. For more complex waveguide geometries the resonances of the atom-surface interaction as a function of the  waveguide  parameters can be more difficult to predict. 

The paper is organized as follows:
In Sec.~\ref{sec1} we evaluate the CP-induced clock frequency shifts and in Sec. \ref{seclifetime} we consider other effects, such as atomic loss and heating inside the fiber. The summary is given in Sec. \ref{secsum}.
In Secs.~\ref{sec1} and \ref{seclifetime}, theoretical expressions are given in Gaussian units,
while in Sec. \ref{secsum}, we also use atomic units, as indicated.

\section{ $^1S_0$-$^3P_0$ Clock transition inside the hollow core fiber \label{sec1} }
The probability of BBR induced transitions from the  clock  levels to the other atomic states is small over the typical clock  operation time and can be neglected, because the clock atoms are not in thermal equilibrium with the BBR bath \citep{GorDuc06}.  The CP interaction induced shift of
the  clock frequency has to be found as the difference of free energy shifts of the clock levels~\cite{GorDuc06,EliBuhSch10}:
\begin{eqnarray}
\delta F_a(\mathbf{r})=\frac{4\pi}3\sum\limits_{j}\{n(|\omega_{ja}|)-\Theta(\omega_{aj})[1+2n(|\omega_{aj}|)]\}|\mu_{aj}|^2\nonumber  \\ \times \mathrm{Tr}[\mathrm{Re}\textbf{G}(\mathbf{r},\mathbf{r},|\omega_{ja}|)]-k_B T{\sum\limits_{k=0}^\infty}'\mathrm{Tr}[\textbf{G}(\mathbf{r},\mathbf{r},i\xi_k)]\alpha_{a}(i\xi_k)\label{shift}, \nonumber \\
\end{eqnarray}
where $\xi_k=2\pi k_BT\cdot k/h$ is the $k$-th Matsubara frequency, $T$
is the temperature, $\mu_{aj}$ is the  dipole matrix element of transition $a\rightarrow j$
with transition frequency $\omega_{ja}$,
$\textbf{G}(\mathbf{r},
\mathbf{r},\omega)$ is the classical Green tensor \cite{WylSip84,WylSip85,LiLeo200}  for a given waveguide  geometry, 
$\alpha_{a}$ is the atomic polarizability,
the prime on the Matsubara sum indicates that the $k= 0$ term is to be taken with half weight,
and $\Theta$ is the Heaviside step function.
We do not present the details of the full theoretical framework used in these calculations as it was previously done by other authors, see Refs.  \cite{GorDuc06,EliBuhSch10} and references therein~\footnote{The factor $\xi^2_j/(c^2\epsilon_0)$ 
in Eq.~(2.3) and the factor $\mu_0 \omega_{kn}^2$ in Eq.~(2.7) of Ref.~\protect\cite{EliBuhSch10} are included through the factor  $\omega^2/c^2$, expressed in Gaussian units, appearing in our expression for the Green tensor, see Eq.~\protect(\ref{GF}) below, and accounts for the factor $4\pi$ in our Eq.~\protect(\ref{shift}).}
We approximate the geometry of the fiber inner interface by a dielectric capillary  of a given thickness $d=R_{out}-R_{in}$, see Fig.~\ref{Fig:Setup}(a), where $R_{in}$ and $R_{out}$ are the inner and outer radii of the capillary
and $\rho$ is the radial coordinate of the atom inside the cylinder.
We consider the surface in thermal equilibrium with BBR at given temperature $T=T_S$.  
The mean occupation number $n$ of  photons with energy $\hbar \omega_{ja}$  is given by the Bose-Einstein distribution, $n(\omega_{ja})=(e^{\hbar\omega_{ja}/{k_BT_S}}-1)^{-1}$. Cold atoms prepared in a given state (either  $^1S_0$ or $^3P_0$) are trapped by the red-detuned  standing wave forming the 1D optical lattice inside the fiber, see Fig.~\ref{Fig:Setup}. The ``magic" wavelength of the laser field forming the  lattice is chosen in order to cancel the differential ac Stark shift for the clock transition,  ($\lambda_L=360$ nm \cite{HacMiyPor08}).  The atomic ensemble temperature can be reduced down to  $\sim$  few nK\citep{HacMiyPor08}, decreasing the  trapping potential depth to $U \sim 10 E_R$, where $E_R$ is the recoil energy.  We also assume that the atoms  are trapped in the ground vibrational motional state and are located near the lattice nodes, corresponding to the maxima of lattice field intensity,  along the axis of a cylinder ($\rho=0$), where they are most distant from the dielectric walls.    
\begin{figure}[h]
\begin{center}
\includegraphics*[width=3.2 in]{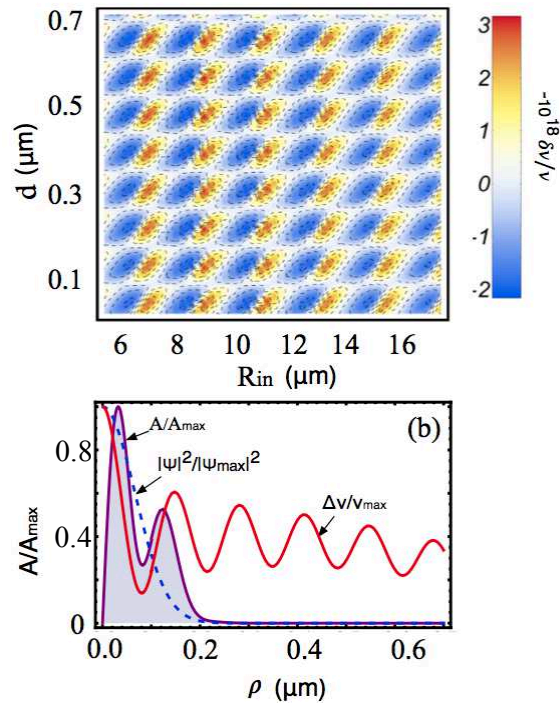}
\end{center}
\caption
{(a) The $^{199}\mathrm{Hg}$  $^1S_0$-$^3P_0$ clock transition frequency shift due to the resonant part of atomic interaction with the inner surface of a silica capillary as a function of its  thickness $d$ and inner radius $R_{in}$ on the capillary axis $\rho=0$. The rate of the  $^1S_0$-$^3P_0$ transition $A=0.6\, \mathrm{s^{-1}}$  from \cite{TyuFavBil16} was used in our calculations.
(b) The  normalized value of $A=\rho\Delta\nu(\rho)|\psi(\rho)|^2$ (purple solid  line) as a function of atom position $\rho$ inside the capillary at the inner core radius $R_{in}=15$ $\mu m$ and the thickness $d=1$ $\mu m$. The solid red and blue dashed curves show the normalized values of the frequency shift $\Delta\nu/\Delta\nu_{max}$ and the amplitude of the  ground vibrational state wave function for the atom inside the red detuned optical lattice with trapping potential depth $U=180$ $\mu $K.
\label{Fig:decay-shift}}
\end{figure}
\subsection{Resonant atom-surface interaction}
We start our analysis of the $^1S_0$-$^3P_0$ clock transition frequency shift from the first term in Eq.~(\ref{shift}), corresponding to virtual dipole  absorption and emission of thermal photons.  In presence of surface polaritons \citep{GorDuc06} or 
waveguide modes  at the atomic transitions frequencies $\omega_{ja}$  these terms may become resonant \cite{EliBuhSch10}. The dashed lines in Fig.~\ref{Fig:Setup}(b) indicate the virtual transitions contributing to the resonant part of the CP interaction potential. The term arising from virtual emission  $^3P_0\rightsquigarrow$ $^1S_0$ remains finite  even at zero temperature, although it is strongly suppressed  due to the small values of the corresponding dipole matrix elements  for non-zero nuclear spin isotopes (which vanishes for nuclear spin-zero isotopes).  The other contributions come from  virtual absorption of thermal photons via the $^3P_0$-$^3S_1$, $^3P_0$-$^3D_1$,$^1S_0$-$^1P_1$, and $^1S_0$-$^3P_{1,0}$ pathways. Their amplitudes are suppressed exponentially with decreasing surface temperature. 

Fig.~\ref{Fig:decay-shift}  shows the computed fractional frequency shift of the $^1S_0$-$^3P_0$ clock transition on the capillary axis ($\rho=0$) caused by  the resonant part of the CP interaction potential  at the surface temperature $T_S=293 $~K.  One can see the resonant structure in its dependence on the capillary thickness parameter $d$ and inner radius $R_{in}$.  The effect ceases (on the capillary axis) as the capillary inner radii grows.
For the vacuum-silica-air interface \cite{GraFesHof16} the cavity effect is small as compared to the case of highly reflective capillary material, Fig.~\ref{Fig:Nonres}. Also, there are  no resonances in the relative permitivity $\varepsilon_{\mathrm{silica}}$  corresponding to transitions from $^1S_0$ or $^3P_0$ states in divalent atoms.  The upper limit on the  resonant CP interaction-induced clock transition frequency shift in Hg atoms on the capillary axis ($\rho=0$), at surface temperature $T_S=293$~K and inner core radius $R_{in}=15 \;\mu \mathrm{m}$ is $|\delta\nu/\nu_{\mathrm{Hg}}|\sim 3 \times 10^{-18}$.  It can be further reduced by choosing the optimal values of waveguide inner radius and thickness. For $^{87}\mathrm{Sr}$ atoms the rate of transition $^{3}P_0$-$^1S_0$, most contributing to the resonant CP interaction at the room temperature of the surface, is much lower than for  $^{199}\mathrm{Hg}$. Therefore we do not consider here the resonant CP interaction induced frequency shifts in Sr atom.

\begin{figure}[h]
\begin{center}
\includegraphics*[width=3.5in]{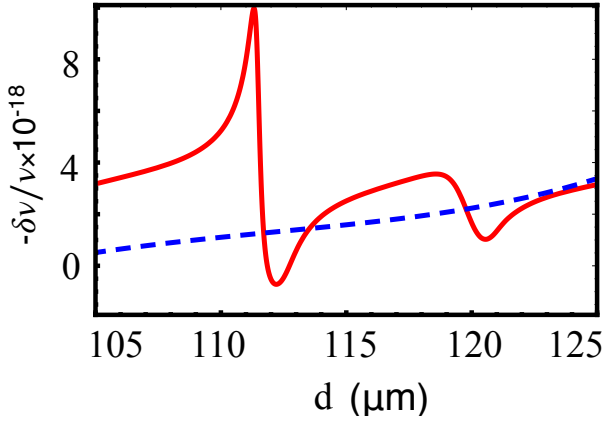}
\end{center}
\caption
{The $^{199}\mathrm{Hg}$  $^1S_0$-$^3P_0$ clock transition frequency shift due to the resonant part of the atomic interaction with the inner surface of a silica capillary as a function of its  thickness $d$ at the fixed value of the  inner radius $R_{in}=15 \;\mu \mathrm{m}$. The atoms are on the capillary axis $\rho=0$, at two different values of relative permittivity of capillary material: $\epsilon_r=2.45$ (blue dashed) and $\epsilon_r=11$ (red solid). \label{Fig:difeps}  }
\end{figure}
The dependence of the resonant CP interaction on the waveguide geometry can be derived from the Green function.   The  Green function is given by \citep{LiLeo200}:
\begin{eqnarray}
\mathrm{Tr}[\mathrm{Re}\textbf{G}(\mathbf{r},\mathbf{r},\omega)]=\frac{i\omega^2}{2\pi c^2}\int\limits_0^{\infty}{dq}{\sum\limits_{m=0}^\infty}' \left\{    \left( r_M+r_N\frac{q^2}{k^2}\right)\right.\nonumber\\ \times \left.\left[ 
\frac{m^2}{\eta^2\rho^2}J_m^2(\eta\rho) +{{J_m'}^2}(\eta\rho)      \right] +r_N\frac{\eta^2}{k^2}J_m^2(\eta\rho)     \right\},\label{GF}
\end{eqnarray}
where $r_{M,N}(n,q)$ are functions of the frequency $\omega$ and the surface parameters: $R_{in}$, $d$, $\varepsilon$.  The resonances of the $r_{M,N}$ coefficients at $q=0$ determine the true resonances of the  Green function. 
On the axis ($\rho=0$) the resonant waveguide mode at the given  frequency $\omega$ can be suppressed when minimizing the corresponding sum   $\mathrm{Im}[r_N(0,0)+r_M(1,0)/2]$. 

\subsection{Nonresonant Casimir-Polder interaction potential}
\begin{figure}[h]
\begin{center}
\includegraphics*[width=3.5in]{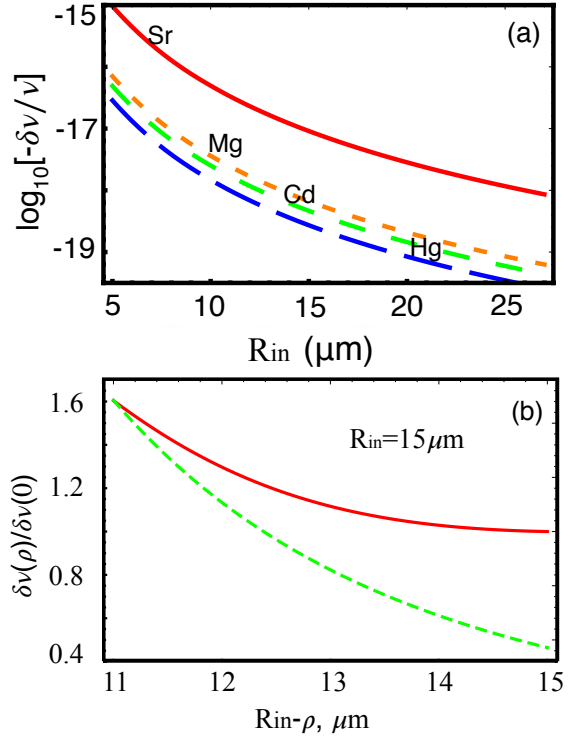}
\end{center}
\caption
{(a) Frequency shift of the $^1S_0$-$^3P_0$ clock transitions in Hg (blue long-dashed), Cd (green medium-dashed), Mg (orange short-dashed) and in Sr (red solid) atoms due to the (dominant)  non-resonant part of the Casimir-Polder interaction potential. Clock atom  are assumed to be located on the capillary axis. The capillary of thickness $d=1\;\mu \mathrm{m}$  has temperature $T_S=77\, \mathrm{K}$.
(b) Frequency shift of the $^1S_0$-$^3P_0$ clock transition in Hg atom  as a function of the distance $R_{in}-\mathrm{\rho}$ between the atom and the capillary inner surface. The capillary has the same parameters as in (a) with fixed inner radius $R_{in} = 15 \, \mathrm{\mu }m$. The green dashed line shows the value $\delta\nu(\rho=11\mathrm{\mu} m) \times (\frac{11\mathrm{\mu}m}{R_{in}-\rho})^4$ expected for CP interaction in planar geometry. \label{Fig:Nonres}   }
\end{figure}

The second term in Eq.~(\ref{shift}) corresponds to non-resonant quantum fluctuations of the atomic dipole. Away from  resonances  and at low surface temperatures this term is the primary contributor to the $^1S_0$ - $^3P_0$ transition frequency shift in divalent atoms.  

Fig.~\ref{Fig:Nonres} (a) shows the calculated  clock  transition frequency shift 
for the Hg, Cd, Mg and Sr atoms on the silica capillary  axis ($\rho=0$) at surface temperature $T_S=77$~K.
In evaluating the shifts we used the dynamic electric
polarizabilities from Refs.~\cite{DerObrDzu09,DerPorBab10,Dzu16}.
The Hg atom is least sensitive to the nonresonant atom surface interaction due to its relatively low differential static polarizability $\delta\alpha=\alpha_{^3P_0}-\alpha_{^1S_0}$ compared to the other divalent atoms (see Ref.~\cite{DerObrDzu09} for extensive discussion of this point).  

Fig.~\ref{Fig:Nonres} (b) shows the calculated  clock CP-induced  transition frequency shift  for  Hg  as a function of distance $R_{in}-\mathrm{\rho}$ between the atom and waveguide surface at the fixed value of inner waveguide radius $R_{in}=15\mathrm{\mu} m$.  $\rho$ is the atomic position in cylindrical coordinates with $\hat{z}$-axis placed at the waveguide axial axis. For a finite-size distribution of atomic motional wavefunction over the radial trapping potential, this dependance will lead to broadening of clock transition line.

\begin{figure}[h]
\begin{center}
\includegraphics*[width=4.2in]{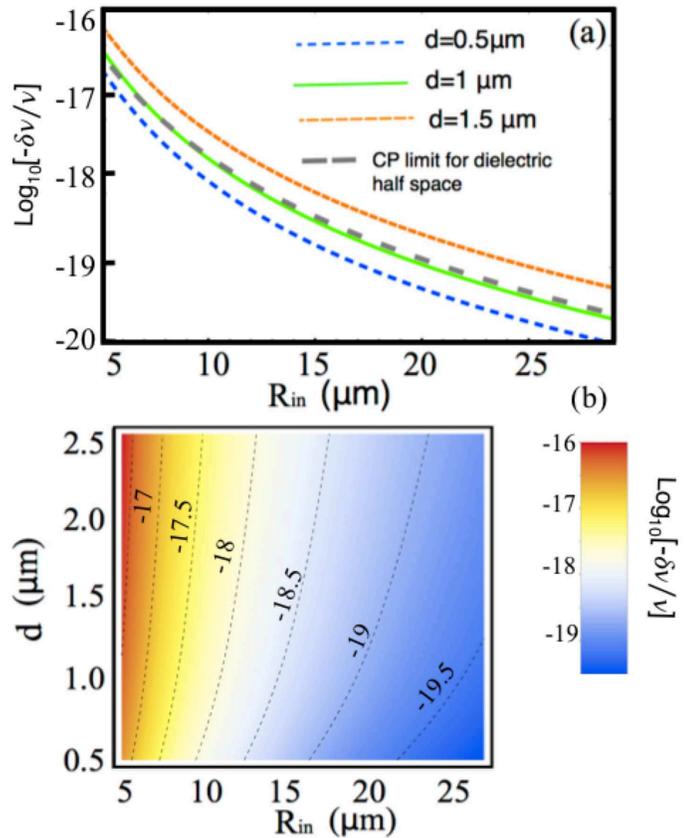}
\end{center}
\caption
{(a) The frequency shift of the $^1S_0$-$^3P_0$ transition in $\mathrm{Hg}$ atoms originating from
the  nonresonant part of CP interaction potential, as a function of the capillary inner radius $R_{in}$ at different  values of its thickness $d$. The gray dashed line shows the CP interaction limit for the dielectric half space interface. (b) The dependence of the $^1S_0$-$^3P_0$ transition frequency shift (non resonant part) in $\mathrm{Hg}$ atom on the waveguide inner radius $R_{in}$ and thickness $d$.  
\label{Fig:HgNonres}}
\end{figure}

Fig.~\ref{Fig:HgNonres} (a) illustrates the  $\mathrm{Hg}$  $^1S_0$-$^3P_0$ transition frequency shift as a function of capillary inner radius at different values of the thickness parameter $d$. The gray dashed line shows the CP frequency shift for an atom interacting with the plane dielectric interface at the distance $z=R_{in}$,  $\delta\nu_{ab}^{CP}(z) =-(3/8\pi) c \alpha_{ab}(0) z^{-4}$  \cite{CasPol48}, where $\alpha_{ab}(0)$ is the differential static polarizability of the atomic levels $a$ and $b$. As the thickness of the capillary grows, the atoms on the  axis are subject to stronger perturbation of their energy levels compared to the case of an atom near the flat dielectric interface. So, for example, at  thickness $d=0.5\, \mathrm{\mu}m$ and  inner radius $15\, \mathrm{\mu}m$ the atomic transition frequency shift is about 2.17 times less than the corresponding value  for the flat interface: $\delta\nu_{ab}^{CP}(z)/\delta\nu_{ab}(z)=2.17$. At the same inner radius value and the thickness $d = 2.5\, \mathrm{\mu}m$ the atomic transition frequency shift is  about 1.92  times larger than for the flat interface: $\delta\nu_{ab}(z)/\delta\nu_{ab}^{CP}(z)=1.92$.  As a comparison, in \cite{AfaMin10} the electrostatic approach was used to determine the van der Waals interaction of the atom with the inner surface of dielectric cylinder ($d\rightarrow\infty$) . The  interaction potential, as a function of atom distance $x$ from the  surface, was presented there as $U(x)=- \mu C_3/x^3$, with the factor $\mu(\rho)$ depending on the distance $\rho$ of the atom from the center of a cylinder. It was found that on the axis of a cylinder the interaction potential felt by the atom is 4 times stronger  ($\mu(0)=4$)  than for the atom near the dielectric  plane surface at the distance equal to the cylinder radii. 

Fig.~\ref{Fig:HgNonres} (b) shows the nonresonant CP interaction induced  $^1S_0$-$^3P_0$ clock transition frequency shift in $^{199}\mathrm{Hg}$ atoms as a function of the capillary inner radius and thickness.
One can see the steady frequency shift growth (absolute value) when increasing the capillary thickness and reducing  the inner radius of the waveguide.

\section{ Lifetime of trapped atoms and other systematic effects \label{seclifetime}}
In this section we further discuss the systematic effects limiting the lifetime of atoms in optical lattice inside the fiber and therefore the stability of fiber
clock. We compare different frequency shifts for the  $^{199}\mathrm{Hg}$ atom, which is the main focus of this paper, with the $\mathrm{Sr}$ atom, used in the most accurate to this date optical lattice clock \cite{BloNicWil14}. 

One of the detrimental effects comes from the collisions of the atoms with the residual buffer gas molecules. Compared to the macroscopic vacuum chamber, achieving  high vacuum in the hollow core  fiber is more  challenging due to the smallness of inner core.   Collisions of cold atoms with the background gas molecules inside the fiber lead to their heating and eventual escape from the optical dipole trap.  Considering that the average kinetic energy of the incident buffer gas molecule $E_{\mathrm{bg}}$ is much larger than that of the trapped atoms and that the scattering process (inside  the shallow optical lattice) happens at relatively large internuclear distances $R$ (so that $|V_{\mathrm{col}}(R)=-C_6/R^6 | \ll E_{\mathrm{bg}}$), one can consider the energy exchange between the two particles using the impulse approximation \cite{John87}. The resulting loss rate is given by $\gamma_{\mathrm{loss}}=2\sqrt{\pi}n\cdot \left(\frac{2k_BT_b}M\right)^{1/3}\left( \frac{15\pi C_6}{8\sqrt{2mU}}\right)^{1/3} \Gamma\left(\frac{11}6\right)$, where $n$ is the buffer gas number density, $M$ and $m$ are, respectively, the masses of the buffer gas molecule and trapped atom, $U$ is the  trapping potential depth, and $T_b$ is the temperature of the buffer gas. The residual buffer gas pressure in the experiments with cold  atoms inside the hollow core fiber \cite{OkaTakBen14, BajHofPey11} is  $P_{bg}\sim 10^{-6}\, \mathrm{Pa}$.  The collisions of cold Hg atoms with molecular nitrogen  $\mathrm{N}_2$ at buffer gas pressure $P_{\mathrm{N}_2}=10^{-6}\, \mathrm{Pa}$, temperature $T_{\mathrm{N}_2}=293$ K and trapping potential depth $U=180$ $\mu $K results in a Hg loss rate $\gamma_{\mathrm{loss}}=0.53$ $1/s$. This corresponds to the half of the atoms escaping the optical lattice trap after 1 s.
In our calculations we used the van der Waals interaction constants $C_6$ given in \cite{Mar32}. The atom loss limits the interrogation time of the atomic ensemble $\tau_{int}\sim 1/\gamma_{\mathrm{loss}}$, and the maximum number of atoms which can be loaded into the optical lattice inside the waveguide. Stronger lattice field intensities can be used during the guiding of the atoms along the fiber.  After the atoms are distributed over the fiber length they can be decelerated and further cooled. The intensity of the lattice field can be then reduced. The lower residual gas pressure is required for further improvement of the clock effective operational time and accuracy. After interrogating the atomic ensemble over the  time $\tau\sim1/\gamma_{\mathrm{loss}}$ the lost atoms need to be evacuated from the fiber and the loading process to be repeated. The residual buffer gas density  inside the  fiber can be further reduced using  light induced desorption  \cite{AtuDanMik14} together with ultra high vacuum pumping techniques.

The collisional heating  of the trapped atoms can be estimated as  $\frac{dT_{\mathrm{Hg}}}{dt}=2\pi\frac{M^2m}{k_B(M+m)^2}v_r^3n\cdot \int\limits_{\pi-2\theta_{\mathrm{min}}}^{\pi}  \overline{\sigma}(\chi)\cos\frac{\chi}2d\chi$, where $v_r$ is the most probable speed of the  buffer gas molecule, $\overline{\sigma}(\chi)=\sigma(\chi)|4\sin\frac\chi2|$, $\sigma(\chi)$ \cite{John87}  is the differential collision cross section, $\chi=\pi-2\theta$,  $\theta_{\mathrm{min}}=\cos^{-1}(1-\frac{Um}{\mu^2v_r^2})$ is the minimum scattering angle corresponding to the escape of initially trapped  atom from the optical lattice, and $\mu=\frac{mM}{m+M}$ is the reduced mass.  Using the impulse approximation \cite{John87} one can obtain: $\overline{\sigma}(\chi)=\frac1{12}\left(\frac{5\pi(m+M)C_6}{4mE_{bg}\sin^2(\frac\chi2)}\right)^{\frac13}$. The corresponding heating rate is $\frac{dT_{\mathrm{Hg}}}{dt}=\frac{\pi}5\frac{mM^2}{k_B(m+M)^2}\cdot n v_r^3\left(\frac{5\pi(m+M)C_6}{4mE_{bg}}\right)^{\frac13} \left(1-(1-\frac{Um}{\mu^2v_r^2})^{\frac53}\right)$. At  the previously specified buffer  gas  parameters the calculated collisional heating rate is $\frac{dT_{\mathrm{Hg}}}{dt}=5.6$ $ \mu \, \mathrm{K/s}$. To calculate the  heating and loss rates for Sr atom we estimated the van der Waals coefficient as $C_{6,\mathrm{Sr}}=\frac32\frac{I_{\mathrm{Sr}}I_{\mathrm{bg}}}{I_{\mathrm{Sr}}+I_{\mathrm{bg}}}\frac{\alpha_{\mathrm{Sr}}\alpha_{\mathrm{bg}}}{(4\pi\epsilon_0)^2}$, where $\alpha_{\mathrm{Sr,bg}}$ and $I_{\mathrm{Sr,bg}}$ are the polarizabilities and ionization potentials of the colliding Sr atom and buffer gas molecule. The ionization energy of molecular nitrogen is $I_{\mathrm{N}_2}=1503\,\mathrm{kJ\,mol^{-1}}$. The polarizability~\cite{SpeMey99} $\alpha_{\mathrm{N}_2}=11.74$  a.u. The corresponding Sr loss rate is $\gamma_{\mathrm{Sr,loss}}=1.03$ $1/s$, $\frac{dT_{\mathrm{Sr}}}{dt}=9.87\, \mu \mathrm{K/s}$.   The additional laser field could be used to cool the atoms during the  ``dark periods" between the  subsequent interrogations. 

The rate of spontaneous photon scattering from the lattice field ($ \lambda_L=360$ nm) by Hg atoms trapped in the ground  $^1S_0$ state is given by $\gamma_{\mathrm{scat}}\sim\frac{\Omega_{\mathrm{Rabi}}^2}{4\Delta^2}$ \cite{MilCliHei93}, where $\Omega_{\mathrm{Rabi}}$ is the Rabi frequency of the  $^1S_0$-$^3P_1$ transition, $\Delta=\omega_{\mathrm{res}}-\omega_{\mathrm{lattice}}$. Considering the large detuning between the frequencies of the lattice field and the resonant frequency of  the $^1S_0$-$^3P_1$ transition, such that $\Delta\gg\Omega_{\mathrm{Rabi}}$, the heating due to the spontaneous scattering can be neglected. 

Another source of heating comes from the lattice intensity noise \cite{SavOHaTho97}. Corresponding heating rate is given by $\frac{dT_{\mathrm{noise}}}{dt}=\sum\limits_{N}P_{N}(N+\frac12)\hbar\Omega_{\mathrm{trap}}\cdot \frac{\Omega_{\mathrm{trap}}^2}{4}\cdot S_\epsilon(2\nu_{\mathrm{trap}})$, where $S_\epsilon(2\nu_{\mathrm{trap}})$ is the fractional intensity noise power spectrum, evaluated at the twice the trapping frequency $\nu_{\mathrm{trap}}$, $N$ is the vibrational state of the trapped atom inside the lattice, and $P_N$ is the population of the given vibrational state $N$.
Using the polarizability values given in \cite{KatOvsMar15}  the lattice intensity noise produced heating rate of $\mathrm{Hg}$ atoms in the ground vibrational state can be estimated as: $\frac{dT_{\mathrm{Hg}}^{\mathrm{noise}}}{dt}=534 (\frac{I_0}{\mathrm{kW/cm^2}})^{\frac32}S_\epsilon(2\times 13.1\sqrt{\frac{I_0}{\mathrm{kW/cm^2}}} \,\mathrm{kHz})\mathrm{K\,s}^{-1}$, where $I_0$ is the lattice field intensity.  For the lattice  potential depth  $U$ $\sim$ $10E_{\mathrm{R,Hg}}=3.67\,\mu \mathrm{K}$ ,  corresponding heating rate is $\frac{dT_{\mathrm{Hg}}^{\mathrm{noise}}}{dt}\approx2.59\cdot 10^4 S_\epsilon(94 \mathrm{kHz})\, \mathrm{K }s^{-1}$. As an example, we take the  fractional intensity noise power spectrum of an argon ion laser \cite{SavOHaTho97} often used to pump the Ti:sapphire and dye lasers \cite{ScoLanKol01}. The resulting heating rate is $\frac{dT_{\mathrm{Hg}}^{\mathrm{noise}}}{dt}\approx 25.9\, \mathrm{nK }s^{-1} $. For Sr atoms the estimates are: $\frac{dT_{\mathrm{Sr}}^{\mathrm{noise}}}{dt}=3724 (\frac{I_0}{\mathrm{kW/cm^2}})^{\frac32}S_\epsilon(2\times 25.05\sqrt{\frac{I_0}{\mathrm{kW/cm^2}}} \,\mathrm{kHz})\mathrm{K\,s}^{-1}$. For the same fractional intensity noise spectrum and the  potential depth  $U$ $\sim$ $10E_{\mathrm{R,Sr}}=1.65\,\mu \mathrm{K}$, one has $\frac{dT_{\mathrm{Sr}}^{\mathrm{noise}}}{dt}\approx 180\, \mathrm{nK }s^{-1} $. 

The residual birefringence of the fiber causes the non-uniformity of the polarization along the lattice. This results in additional clock transition frequency uncertainty. The $^{87}$Sr clock transition frequency shift due to contribution of the vector and tensor polarizabilities  was presented in  \cite{WesLodLor11}  as $\Delta\nu_{v,t}=(0.22 \mathrm{Hz}\cdot m_F\xi (\mathbf{e_k}\cdot \mathbf{e_B})-0.0577\,\textrm{mHz}\cdot\beta)U/E_R$,
 where  $U$ is the lattice potential depth, $E_R$ is the recoil energy, $\mathbf{e_k}$,$\mathbf{e_B}$ are the unitary vectors along the lattice wave vector and the quantization axis,  $\beta=(3|\mathbf{e}\cdot \mathbf{e_B}|^2-1)[3m_F^2-F(F+1)]$, $\mathbf{e}$ is the complex polarization vector, and $\xi$ is the degree of the ellipticity of the lattice field. Taking $(\mathbf{e_k}\cdot \mathbf{e_B})=1$, $\Delta\xi\sim\frac{\pi\Delta n}{\lambda_L}\cdot L\ll1$, where $\Delta n$ is the difference of the refractive indexes for two orthogonal polarizations and $L$ is the length of the atomic cloud inside the fiber, one can estimate the vector polarizability induced frequency shift uncertainty as $\delta\nu_{v,m_F}\sim\frac{\pi}{\lambda_L}m_F\Delta n L\cdot 0.22 \,\mathrm{Hz}$. For $m_F=9/2$, $\Delta n\sim 10^{-7}$ \cite{OkaTakBen14}, $\delta\nu_{v\frac92}=3.1\,\mathrm{Hz}\cdot\frac{L}{\lambda}\frac{U}{E_R}\cdot 10^{-7}$. For $L=8.13$  cm, $U=10$ $E_R$, $\delta\nu_{v\frac92}=0.3\,\mathrm{Hz}$. The vector light-shifts of the components $m_F=\pm\frac92$ have the opposite signs. Zeeman shift and vector-light shift cancellation techniques have been developed in \cite{TakHonHig06} based on the averaging of frequency measurements for two transitions ($^1S_0, F=\frac92,m_F=\pm\frac92)- (^3P_0, F=\frac92, m_F=\pm\frac92$).
 For the same parameters $\Delta n$, $L$, $U$, and $(\mathbf{e_k}\cdot \mathbf{e_B})=0$, $(\mathbf{e}\cdot \mathbf{e_B})=1$, $\delta\nu_{t\frac92}=41.5\,\mathrm{mHz}$.  The dipole polarizabilities of the $^{199}\mathrm{Hg}$ atom at the magic wavelength $\lambda_{L}=360\, \mathrm{nm}$ are lower than corresponding values for the Sr atom  \cite{KatOvsMar15}. Therefore the vector and tensor frequency shift are not exceeding those for the $^{87}\mathrm{Sr}$ atom. Although for bosonic isotopes there are no tensor or vector shifts, the polarization instability along the axis may lead to the multipolar effects induced frequency shifts \cite{TaiYudOvs08,KatHasIli09}, which are  however much weaker than the electric-dipole interaction \cite{KatOvsMar15}. One could consider using the circularly polarized optical lattice for the atomic clocks with bosonic isotopes.

\section{summary \label{secsum}}

We have studied the feasibility of the optical lattice clock based on the ultra narrow $^1S_0$-$^3P_0$ transition in Hg and other divalent atoms optically trapped inside the micron-scale hollow core waveguide. 
The  effect of the atom-surface interaction on the clock transition frequency inaccuracy at non-zero surface temperature has been evaluated.  For an ensemble of cold atoms, with the temperature $T_a\sim \mu \mathrm{K}$, optically trapped  on the axis of a silica capillary  (with surface temperature  $T_S=293\;\mathrm{K}$), the main contribution to the surface induced $^1S_0$-$^3P_0$ transition frequency shift comes from the nonresonant part of Casimir-Polder interaction potential.   
This contribution is substantially suppressed for Hg and Cd atoms compared to the other divalent atoms due to 
their relatively low differential static polarizability. For example, at the inner capillary radius $R_{in}=15\,\mathrm{\mu m}$  and the capillary thickness $d=1\,\mathrm{\mu m}$, the CP interaction induced $^1S_0$-$^3P_0$ transition frequency shift in $^{199}\mathrm{Hg}$ atoms is $\delta\nu_\mathrm{Hg}/\nu_\mathrm{Hg}=0.03\delta\nu_\mathrm{Sr}/\nu_\mathrm{Sr}=0.27\times10^{-18}$. 
For the silica capillary waveguide $\epsilon_r=2.45$,  the calculated upper limits on the $^1S_0$-$^3P_0$  transition frequency shift  caused by the resonant atom-waveguide coupling effects are below $3\cdot10^{-18}$.    In general this shift and the natural linewidth broadening (Purcell effect)  can be cotrolled by proper choice of geometric parameters of the waveguide.
One could consider the possibility to compensate  the nonresonant  part  of Casimir-Polder interaction  on the core axis in combination with its resonant part at certain   geometries of the waveguide. 
Additional effects may appear in case of the resonances in dielectric constants of waveguide material.  

We also  estimated the atom loss and heating due to the collisions with the buffer gas particles, lattice intensity noise induced heating, spontaneous photon scattering and the residual birefringence induced frequency shifts. To fully realize the potential of the fiber based clock, one needs to solve the problem of the residual buffer gas pressure limiting the interrogation time of the atomic ensemble and the number of atoms which can be loaded into the fiber before they escape the trapping potential.

We would like to thank K. Gibble and H. Katori for motivating discussions. This work was supported in part by the U.S. National Science Foundation   grant PHY-1607396.

 \end{document}